# EFFICIENTLY SECURE BROADCASTING IN 5G WIRELESS FOG-BASED-FRONTHAUL NETWORKS

Nabeel I. Sulieman and Richard D. Gitlin

Department of Electrical Engineering, University of South Florida
Tampa, Florida 33620, USA

## ABSTRACT

*Enhanced Diversity and Network Coding (eDC-NC), the synergistic combination of Diversity and modified Triangular Network Coding, was introduced recently to provide efficient and ultra-reliable networking with near-instantaneous fault recovery. In this paper it is shown that eDC-NC technology can efficiently and securely broadcast messages in 5G wireless fog-computing-based Radio Access Networks (F-RAN). In particular, this work is directed towards demonstrating the ability of eDC-NC technology to more efficiently provide secure messages broadcasting than standardized methods such as Secure Multicasting using Secret (Shared) Key Cryptography, such that the adversary has no ability to acquire information even if they wiretap the entire F-RAN network (except of course the source and destination nodes). Our results show that using secure eDC-NC technology in F-RAN fronthaul network will enhance secure broadcasting and provide ultra-reliability networking, near-instantaneous fault recovery, and retain the throughput benefits of Network Coding.*

## 1. INTRODUCTION

Many wired and wireless networks securely broadcast/multicast messages to a group of receivers. Several studies [1]-[4] have theoretically analyzed the ability of Network Coding [5] to provide secure broadcasting/multicasting in any wired or wireless network. These papers assume that the number of channels that the eavesdropper can wiretap is equal to or less than the number of tolerated wiretapped channels in the network, which is a design parameter [1]-[4]. However, this assumption is not valid in real networks, as the eavesdropper might have the ability to wiretap the entire network of channels. To overcome this vulnerability, Secret (Shared) Key Cryptography and Network Coding is utilized to provide efficient, secure messages broadcasting. In this study it is shown that Enhanced Diversity and Network Coding (eDC-NC) [6]-[7], which is the synergistic combination of Diversity Coding [8] and modified Triangular Network Coding [9], can efficiently and securely broadcast messages in 5G wireless fog-computing-based Radio Access Networks (F-RANs). F-RANs are an alternative network architecture to Cloud Radio Access Networks (C-RANs) [10]-[12] that are under consideration for 5G networks, where the centralized processing in the baseband unit (BBU) of the C-RAN are replaced by edge nodes with the ability to control, process and store data, and communicate with each other [10]-[12].

It is shown in [6]-[7] that eDC-NC can simultaneously improve the network reliability, reduce computational complexity, enable extremely fast recovery from simultaneous multiple link/node failures, and retain the throughput improvement of Network Coding for broadcasting/multicasting applications of a F-RAN wireless fronthaul networks.

The contribution of this paper is demonstrating the ability of eDC-NC technology to more efficiently provide secure messages broadcasting than standardized methods such as Secure Multicasting [13], such that the adversary cannot acquire any information even if they can wiretap the entire F-RAN network (except of course the source and destination nodes). In this way, eDC-NC will enhance secure broadcasting and provide ultra-reliability networking, near-instantaneous fault recovery, and retain the throughput gain of Network Coding.

The structure of this paper is as follows. The system model based on F-RAN wireless fronthaul network is described in Section 2. Section 3 demonstrates the ability of eDC-NC coding to provide secure broadcasting/multicasting followed by the application of secure eDC-NC to F-RAN wireless fronthaul networks in Section 4 and its efficiency analysis in Section 5. The paper ends with the concluding remarks in Section 6.

## 2. SYSTEM MODEL

F-RANs were proposed in [10]-[12] to enhance the performance of C-RANs by migrating a significant number of functions to edge devices and substantially upgrading the Remote Radio Heads (RRHs). These functions include controlling, communicating, measuring, managing, and storing and processing of data. The upgraded RRH is called a Fog Access Point (F-AP), and is able to communicate with other F-APs. One of the benefits of this architecture is decreasing latency by performing functionality at the network edge rather than in the core [10]-[12]. There are three layers in the architecture of F-RANs as illustrated in Figure 1 [10]. The network layer contains the BBU pool, centralized storage, and communication and computing cloud. The F-APs represents the access layer. The terminal layer contains Fog-user equipment (F-UE) that access the F-AP [10]-[11]. Adjacent F-APs can be formed into two topologies: a mesh topology or a tree-like topology. Both of these topologies can significantly minimize the degrading effects of capacity-constrained fronthaul links [10].

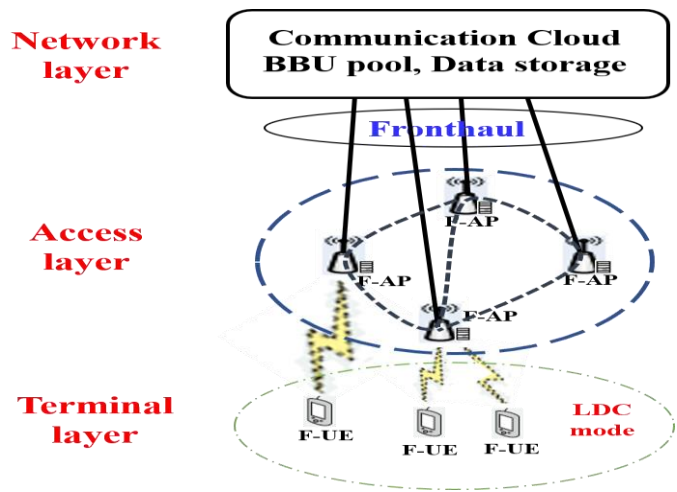

Figure 1.  F-RAN network architecture.

Although there are different transmission modes in a F-RAN, this paper is focused on the Local Distributed Coordination (LDC) mode, as it is the core mode of the F-RAN as shown in Figure 1 [10]. In the LDC, the F-APs communicate with other F-APs to serve the F-UEs. This will decrease the burden on the fronthaul network and quickly suppress interference or transmit the required data to the UE not from the cloud server but from the F-APs [10].

In [7], eDC-NC coding is applied to a wireless sensor network to provide ultra-reliable with near-instant fault recovery and efficient energy consuming system, and here the ability of eDC-NC to provide efficient secure messages broadcasting and apply secure eDC-NC coding to the LDC fronthaul network is demonstrated, where F-APs are connected to each other in a mesh topology as shown in Figure 2. In this paper, the connections are considered to be wireless links and wiretapped by an eavesdropper. To minimize interference and be able to communicate with each other, such F-APs will likely utilize MIMO technology [14].

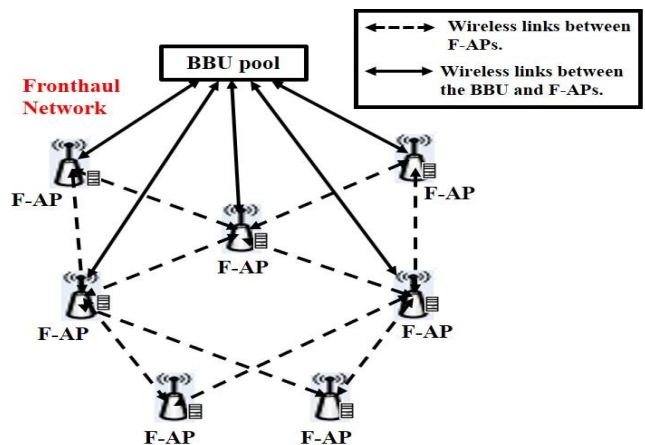

Figure 2.  Example F-RAN fronthaul network with wireless links.

## 3. SECURE ENHANCED DC-NC BROADCASTING NETWORK

In order to illustrate how Secret (Shared) Key Cryptography is used in eDC-NC type networks and to provides secure broadcasting, consider the point-to-multipoint network topology depicted in Figure 3.

Each broadcasting session will have its own secret (shared) session key. The source node (node 1) and the receiver nodes (nodes 5 and 6) will share the broadcasting session key. Controlling the distribution of the keys between the source and the legitimate receivers is a primary issue in any communication network. The IETF Group Domain of Interpretation (GDOI) protocol defined in Request for Comments (RFC-6407) [15] may be used to facilitate connecting the source and the destinations to a key server, where using Public Key Cryptography (PKC) the keys are encrypted and distributed to the members of secure multicast group. The source and the destinations can be authenticated and authorized to form a specific multicast group by the key server such that the shared key is utilized to encrypt and decrypt messages between members of the group [15]. In this way, the broadcasting session (shared) key will be distributed securely to the source and destination nodes.

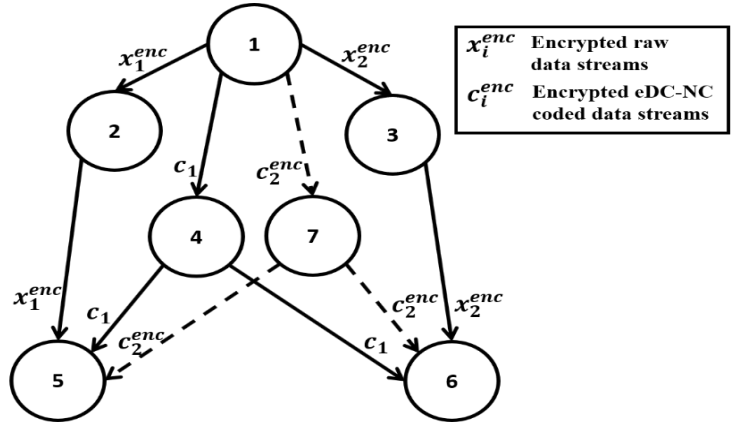

Figure 3. Example wireless network with secure broadcasting via eDC-NC coding that broadcasts two data streams to nodes 5 and 6 and protects each stream from one link/relay node failure. The solid lines represent the wireless links that carry coded data streams and are used to improve network throughput whereas dashed lines represent the wireless links that carry coded data streams and are used to maximize network reliability.

In our example network the source broadcasts two data streams $x_1$ and $x_2$ to destination nodes 5 and 6 using relay nodes 2, 3, 4, and 7.

The system proceeds as follows:

(1) The streams $c_1$ and $c_2$ are created using eDC-NC encoding [6], which will be referred to as eDC-NC coded data streams to distinguish them from the encrypted data streams as follows:

$$c_1 = x_{1,0} \oplus x_{2,1} \quad (1)$$

$$c_2 = x_{1,0} \oplus x_{2,2} \quad (2)$$

where $x_{i,r_i}$ represents the raw data stream $i$ and $r_i$ is the number of redundant "0" bit(s) that are added at the head of raw data stream $i$, $c_i$ represents the eDC-NC coded data stream $i$. Note that the eDC-NC coded data stream, $c_2$, will be encrypted using the shared key at the source node 1 then it will be transmitted. Therefore, it appears in the figure as $c_2^{enc}$.

(2) The source node (node 1) encrypts the streams $x_1$, $x_2$, and $c_2$ using the Secret (Shared) Key Cryptography algorithm.

Node 1 transmits $x_1^{enc}$ and $x_2^{enc}$ to nodes 5 and 6 respectively via relay nodes 2 and 3 respectively. In addition, node 1 transmits $c_1$ in order to realize the throughput gain provided by eDC-NC networking [6]-[7] and $c_2^{enc}$ to be able to tolerate one link/relay node failure [6]-[7] to the destination nodes 5 and 6 via relay nodes 4 and 7 respectively.

(3) At the destination side, for example, node 5 will use the broadcasting session shared key to decrypt the received data streams using the Secret (Shared) Key Cryptography algorithm depending on the following situations:

a. If all data streams are correctly received (by checking the CRC), $c_2^{enc}$ is ignored and $x_1^{enc}$ will be decrypted to find $x_1$. Next, $x_2$ will be recovered by applying $x_1$ to $c_1$ using the eDC-NC decoding algorithm that was explained in detail in [6].

International Journal of Wireless & Mobile Networks (IJWMN) Vol. 10, No. 3, June 2018International Journal of Wireless & Mobile Networks (IJWMN) Vol. 10, No. 3, June 2018

b. If one data stream is either incorrectly received or not received at all (either $c_2^{enc}$ or $x_1^{enc}$), the receiver will decrypt the one that is correct (correct CRC) and then apply it to $c_1$ to obtain $x_2$ similarly to step a.

c. If $c_1$ fails, the receiver will decrypt both $c_2^{enc}$ and $x_1^{enc}$ to get $c_2$ and $x_1$ then similarly to step a apply $x_1$ to $c_2$ to obtain $x_2$.

It is worth noting that the intermediate (relay) nodes will not need to decrypt the encrypted raw and eDC-NC coded data streams because they only need to forward them to the destination nodes. Hence, they do not have to be secured. Also, they CANNOT decrypt the encrypted raw and eDC-NC coded data streams because they do not have the broadcasting session secret (shared) key.

Now, suppose that the adversary wiretaps the entire network, he/she will have $x_1^{enc}$, $x_2^{enc}$, $c_1$, and $c_2^{enc}$. Only $c_1$ is not encrypted but it will not disclose any information because it is a XOR combination of raw data streams $x_1$ and $x_2$. The other data streams are encrypted so, the adversary will need to know the broadcasting session secret (shared) key to decrypt them.

In this way, as long as the adversary does not possess the broadcasting session key, it will not be able to get any information even if he/she wiretaps the entire network (except of course the source and destination nodes).

## 4. APPLYING SECURE ENHANCED DC-NC CODING TO F-RANS

Enhancing the security (privacy) of transmitted information in 5G wireless F-RAN fronthaul networks is critical due to the vulnerability of wireless links. In addition, although F-APs have specific resources, but these should be used efficiently. Enhanced DC-NC technology was recently utilized to improve the efficiency and reliability of 5G wireless F-RAN fronthaul networks and provide rapid recovery time from multiple simultaneous link/node failures while retaining the throughput enhancement feature of Network Coding for broadcasting applications [6]. Here, secure eDC-NC technology is applied to Local Distributed Coordination (LDC) transmission mode in an F-RAN network to achieve the above benefits and to efficiently improve its security, as depicted in Figure 4.

In this example, F-APs are connected to each other in a mesh topology by bi-directional wireless links. Here, a point-to-multipoint network topology models the application of securely broadcasting three data streams $x_1$, $x_2$, and $x_3$ from one F-AP (F-AP1) to three F-APs (F-APs 6, 7, and 8) using intermediate relay nodes F-APs 2, 3, 4, 5, and 9. As mentioned in Section 3, each broadcasting session will have its own secret (shared) session key. The source node (F-AP1) and the receiver nodes (F-APs 6, 7, and 8) will share the broadcasting session key using the GDOI protocol.

The system proceeds as follows:

(1) The coded data streams $c_1$, $c_2$, and $c_3$ are created using eDC-NC encoding [6] as follows:

$$c_1 = x_{1,0} \oplus x_{2,1} \oplus x_{3,2} \quad (3)$$

$$c_2 = x_{1,0} \oplus x_{2,2} \oplus x_{3,1} \quad (4)$$

$$c_3 = x_{1,0} \oplus x_{2,3} \oplus x_{3,5} \quad (5)$$

Note that the eDC-NC coded data stream ($c_3$) will be encrypted using the shared key at F-AP1 then it will be transmitted. Therefore, it appears in the figure as $c_3^{enc}$.

55

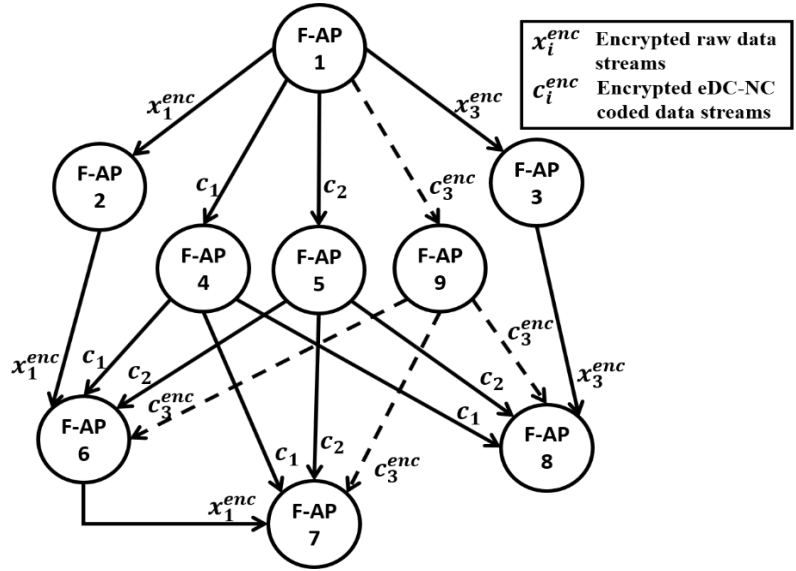

Figure 4. Example wireless fronthaul Fog-RAN network with secure eDC-NC coding that broadcasts three data streams to F-AP6, F-AP7, and F-AP8 and protects each stream from one link failure. The solid lines represent the links that carry coded data streams and are used to improve network throughput whereas dashed lines represent the links that carry coded data streams and are used to maximize network reliability.

(2) F-AP1 encrypts the streams $x_1$, $x_3$, and $c_3$ using the Secret (Shared) Key Cryptography algorithm. Note that data stream $x_2$ does not need to be encrypted because there is no need to transmit it in a separate link. However, it can be recovered from the coded data streams at the destination F-APs.

F-AP1 transmits $x_1^{enc}$ and $x_3^{enc}$ to F-AP6 and F-AP8 respectively via relays F-APs 2 and 3 respectively. In addition, $c_1$ and $c_2$ are transmitted to realize the throughput gain provided by eDC-NC [6]-[7] and $c_3^{enc}$ to be able to tolerate one link/relay F-AP failure via eDC-NC [6]-[7] to the destination F-APs 6, 7, and 8 via relays F-APs 4, 5, and 9 respectively.

(3) At the destination side, for example, F-AP6 will use the broadcasting session shared key to decrypt the arrived data streams via Secret (Shared) Key Cryptography algorithm based on the following situations:

a. If all data streams are correctly received (by checking the CRC), $c_3^{enc}$ is ignored and $x_1^{enc}$ will be decrypted to find $x_1$. Next, $x_2$ and $x_3$ will be recovered using the eDC-NC decoding algorithm that was explained in detail in [6].

b. If one data stream is either incorrectly received or not received at all (either $c_3^{enc}$ or $x_1^{enc}$), the receiver will decrypt the one that is correct (by checking the CRC) and then obtain all data streams in similar way of step a.

c. If $c_1$ or $c_2$ fails, the receiver will decrypt both $c_3^{enc}$ and $x_1^{enc}$ to recover $c_3$ and $x_1$ then similarly as in step a obtain $x_2$ and $x_3$.

As it is mentioned in Section 3, the intermediate (relay) F-APs will not need to decrypt the encrypted raw and eDC-NC coded data streams because they only need to forward the streams to the destination F-APs. Also, they CANNOT decrypt the encrypted raw and eDC-NC coded data streams because they do not have the broadcasting session secret (shared) key.

Now, assuming that the eavesdropper wiretaps the entire network, he/she will have $x_1^{enc}$, $x_3^{enc}$, $c_1$, $c_2$ and $c_3^{enc}$. Although $c_1$ and $c_2$ are not encrypted, no information can be disclosed because they are XOR combinations of raw data streams $x_1$, $x_2$, and $x_3$. Other data streams are encrypted so, the eavesdropper will need to know the broadcasting session secret (shared) key to decrypt them.

In this way, as long as the adversary does not have the broadcasting session key, it will not be able to get any information even if he/she wiretaps the entire network (except, of course, the source and destination nodes).

Consequently, secure eDC-NC will efficiently enable secure broadcasting and provide ultra-reliability networking, near-instantaneous fault recovery, and re the throughput gain of Network Coding of 5G wireless F-RAN fronthaul networks.

## 5. EFFICIENCY ANALYSIS

Normally, when standard Secure Multicast [13] is utilized to provide network security, the source has to encrypt all transmitted data streams. However, by applying eDC-NC, the source node does not need to encrypt all the streams as shown in Section 3 and 4. In the example network in Section 3, only three out of four data streams have to be encrypted, namely $x_1$, $x_2$, and $c_2$. The reason for encrypting only three data streams is that the eDC-NC coded data streams $c_1$ and $c_2$ are not plaintext but the mod 2 combination of the raw data streams $x_1$ and $x_2$. So, encrypting one stream ($c_2^{enc}$) in this example network will make recovering the raw data streams impossible without knowledge of the broadcasting session key to decrypt $c_2^{enc}$ and thus be able to recover both raw data streams via the eDC-NC decoding algorithm [6]. In this way, the cost/complexity of encryption will be decreased by 25%. At the receiver, each destination node has only to decrypt two out of three data streams as a maximum as illustrated in case c in Section 3 above. Hence, the cost of decryption will be maximum and for two broadcast data streams, there will be no cost benefits in decryption, (however, for three broadcast data stream, the decryption cost will be decreased by 33.33%) which is referred to as a minimum decryption cost benefits (*Min. DecCB*). However, in some cases, the destination node has only to decrypt one data stream (cases a and b in Section 3) which decreases the decryption cost by 50%. In this case, the cost of decryption will be a minimum and there will be 50% decryption cost benefit, which is referred to as a maximum decryption cost benefits (*Max. DecCB*). Note that the encryption cost benefit will not vary once the system parameters are fixed because the source node always needs to encrypt only two raw data streams and once the number of eDC-NC encoded data stream(s) that will be used to improve the system reliability has been selected (that is, the number of link failures that can be tolerated) the benefit will be determined.

In general, for point-to-multipoint network topology, in order to quantify the security cost benefits, we need to define the following variables:

International Journal of Wireless & Mobile Networks (IJWMN) Vol. 10, No. 3, June 2018$D_{total}$: Number of overall transmitted data streams.

$D_{enc}$: Number of encrypted (raw and eDC-NC coded) data streams.

$D_{raw}$: Number of broadcast raw data streams.

$L_f$: Number of link failures that can be tolerated in the network.

$D_{dec}$: Number of decrypted data streams.

Therefore, the security cost benefit can be calculated as follows:

The encryption cost benefits (*EncCB*) in percentage is:

$$EncCB\ \% = \frac{D_{total} - D_{enc}}{D_{total}} \times 100 \quad (6)$$

where

$$D_{total} = D_{raw} + L_f + 1 \quad (7)$$

and

$$D_{enc} = L_f + 2 \quad (8)$$

Hence,

$$EncCB\ \% = \frac{D_{raw} - 1}{D_{raw} + L_f + 1} \times 100 \quad (9)$$

The decryption cost benefits (*DecCB*) in percentage at each destination node is:

$$DecCB\ \% = \frac{D_{raw} - D_{dec}}{D_{raw}} \times 100 \quad (10)$$

where

$$D_{dec} = \begin{cases} 1 & \text{if destination node has no link failure OR only one encrypted data stream is lost} \\ L_f + 1 & \text{otherwise} \end{cases},$$

Table 1 shows the security cost benefits in percentage for <u>tolerance of one link failure</u> and different numbers of broadcast data streams.

Table 1. The security cost benefits in percentage for one link failure tolerance.

| Number of broadcast data streams | Number of overall data streams | Number of encrypted data streams | *EncCB* (%) | Min. *DecCB* (%) | Max. *DecCB* (%) |
|---|---|---|---|---|---|
| 2 | 4 | 3 | 25 | 0 | 50 |
| 3 | 5 | 3 | 40 | 33.33 | 66.67 |
| 4 | 6 | 3 | 50 | 50 | 75 |
| 5 | 7 | 3 | 57.14 | 60 | 80 |
| 6 | 8 | 3 | 62.5 | 66.67 | 83.3 |

Figures 5 and 6 illustrate the relationship between the encryption and decryption cost benefits percentage respectively and the number of broadcast data streams. Note that by applying Secure Multicast [13], all data streams have to be encrypted at the source and decrypted at the destination nodes. However, Figure 5 and Figure 6 show that by applying Secret (Shared) Key Cryptography to the eDC-NC broadcast networks, there are always security cost benefits (except in the case of minimum decryption cost for broadcasting two data streams). Also, they depict the scalability of eDC-NC by decreasing of the security costs with the increasing the number of broadcast data streams.

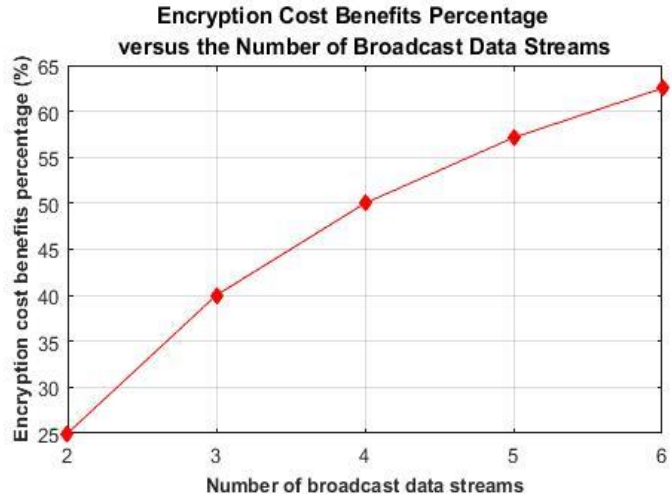

Figure 5.  eDC-NC encryption cost benefits.

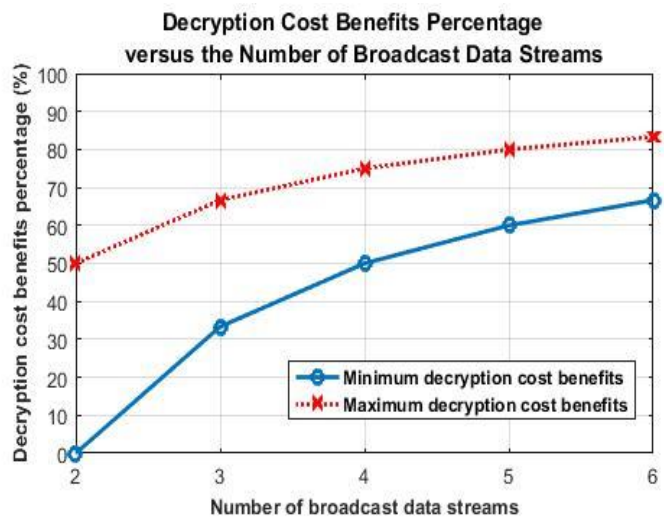

Figure 6.  eDC-NC decryption cost benefits.

However, the security costs will increase with increasing the number of link failures that need to be tolerated as shown in Figure 7 and Figure 8.

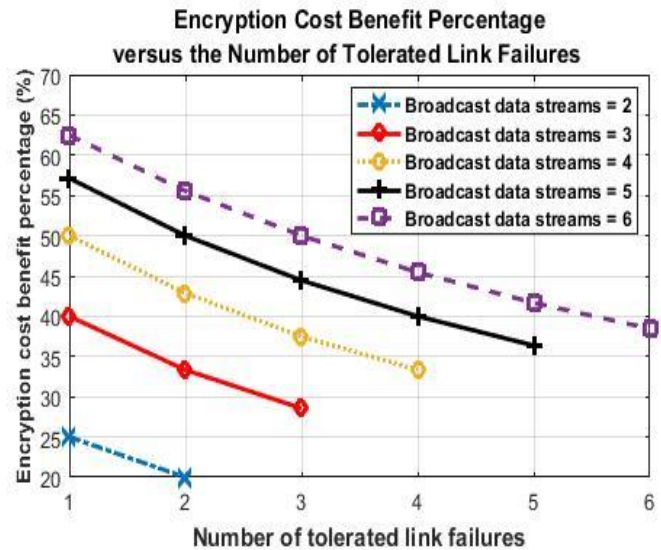

Figure 7. eDC-NC encryption cost benefits for different number of tolerant links.

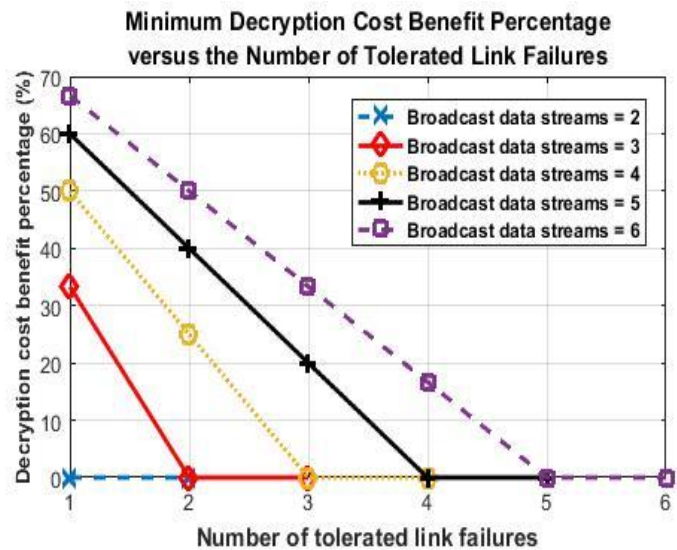

Figure 8. eDC-NC minimum decryption cost benefits for different number of tolerant links.

## 6. CONCLUSIONS

This study presented efficient secure broadcasting via eDC-NC technology for 5G wireless F-RAN fronthaul networks such that the adversary has no ability to acquire any information even if they wiretap the entire fronthaul network (except of course the source and destination F-APs). The security of the broadcasting data streams is obtained with lower security cost compared to that of the standard Secure Multicast protocols. Enhanced secure broadcasting using eDC-NC in F-RAN wireless fronthaul networks provides ultra-reliable communications, near-instantaneous link/node failure recovery, and retains the throughput gains of Network Coding.

## ACKNOWLEDGEMENTS

Nabeel I. Sulieman is supported by the Higher Committee for Education Development in Iraq (HCED-IRAQ).

**AUTHORS**

Nabeel I. Sulieman received his B.S. degree in Electrical Engineering from University of Baghdad, Baghdad – Iraq in 1998, he was one of the ten highest ranked students in the Electrical Engineering Department, and he received his M.S. degree with merit in wireless communications systems from Brunel University, London – UK in 2008. From 2002 until 2014, he worked for Iraqi Telecommunication and Post Company as a technical support engineer. In addition, he worked as an instructor for short technical courses in Higher Institute of Telecommunications-Baghdad-Iraq. Currently, he is a Ph.D. Candidate in Electrical Engineering Department at the University of South Florida in the Innovations in Wireless Information Networking Laboratory (*i*WINLAB) under the supervision of Dr. Richard Gitlin, and his research interests include Diversity Coding, Network Coding, 5G Wireless Fronthaul Networks, Synchronization of Diversity and Network Coding, Software Defined Networking (SDN), and Network Function Virtualization (NFV). 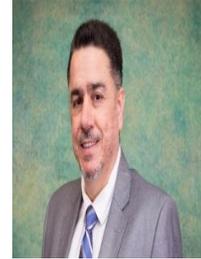

Richard D. Gitlin is a State of Florida 21st Century World Class Scholar, Distinguished University Professor and the Agere Systems Chaired Distinguished Professor of Electrical Engineering at the University of South Florida. He has more than 45 years of leadership in the communications and networking industry. He was at Bell Labs/Lucent Technologies for 32-years performing and leading pioneering research and development in digital communications, broadband networking, and wireless systems. Dr. Gitlin was Senior VP for Communications and Networking Research at Bell Labs and later CTO of Lucent's Data Networking Business Unit. After retiring from Lucent, he was visiting professor of Electrical Engineering at Columbia University, where he supervised several doctoral students and research projects and then Chief Technology Officer of Hammerhead Systems, a venture funded networking company in Silicon Valley. Since joining USF in 2008, he has focused on the intersection of communications with medicine and created an interdisciplinary team that is focused on wireless networking of *in vivo* miniature wirelessly controlled devices to advance minimally invasive surgery and other cyber-physical health care systems. Dr. Gitlin has also directed research on advancing wireless local and 4G and 5G cellular systems by increasing their reliability and capacity. Dr. Gitlin is a member of the National Academy of Engineering (NAE), a Fellow of the IEEE, a Bell Laboratories Fellow, a Charter Fellow of the National Academy of Inventors (NAI), and a member of the Florida Inventors Hall of Fame. He is also a co-recipient of the 2005 Thomas Alva Edison Patent Award and the S.O. Rice prize, has co-authored a text, published ~150 papers and holds 65 patents. 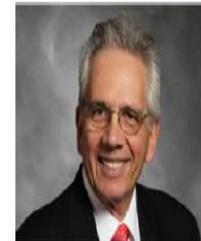